# Drastic reduction of the slow scintillation component in highly luminescent Ce$^{3+}$ and Mg$^{2+}$ doped Lu$_{2.5}$Gd$_{0.5}$Ga$_2$Al$_3$O$_{12}$ garnet powders

LENKA PROUZOVÁ PROCHÁZKOVÁ,$^{1,2*}$ ELIŠKA MÜLLEROVÁ,$^1$ JAN BÁRTA,$^{1,2}$ ESTELLE HOMEYER$^3$, ALENA ZAVADILOVÁ,$^1$ CHRISTOPHE DUJARDIN$^{3,4}$ AND VÁCLAV ČUBA$^1$

$^1$ *Czech Technical University in Prague, Faculty of Nuclear Sciences and Physical Engineering, Břehová 7, 115 19 Czech Republic*
$^2$ *Institute of Physics of the AS CR, v.v.i, Cukrovarnická 10, Prague 6, Czech Republic*
3 *Universite Claude Bernard Lyon 1, ILM, UMR CNRS 5306, 69622 Villeurbanne, France*
4 *Institut Universitaire de France (IUF)*

*\*prochle2@cvut.cz*

**Abstract:** This paper deals with the photochemical preparation of nanomaterials with garnet structure. Ce$^{3+}$ and Mg$^{2+}$ doped Lu$_{2.5}$Gd$_{0.5}$Ga$_2$Al$_3$O$_{12}$ powders were prepared by using UV irradiation of aqueous solutions with low-pressure mercury lamps and subsequent calcination of the solid products. The synthesis was optimized and gives access to a range of doping which is very hard to achieve with single crystal growth from melt. The effect of Ce and Mg concentration on the structural and luminescence properties was studied. Garnets were analyzed using X-ray fluorescence (XRF) and X-ray powder diffraction (XRPD) and their luminescence properties under optical and X-ray photon excitations were investigated. XRF and XRD show that the samples are sufficiently chemically pure and phase-pure and their elemental composition corresponds with expectations. Laser-induced breakdown spectroscopy confirmed that Mg concentrations in Mg-codoped samples are slightly lower than the nominal. Luminescence spectra show typical emission maxima of 5d-4f Ce$^{3+}$ transition and 4f-4f Gd$^{3+}$ transition. The effects of the concentration of Ce$^{3+}$ and Mg$^{2+}$ on the RL intensity, light yields and decays were observed.

## 1. Introduction

Research in the field of multicomponent garnets as highly effective scintillators has been very extensive in the last decades. Band gap engineering of the scintillating properties by the composition adjustment was mentioned since 2011 many times [1,2,3,4,5] and until today, great progress has been made in growing multicomponent garnet crystals [6].

Garnet-based materials provide many advantages for detection of ionizing radiation, such as chemical, mechanical and radiation stability [7,8], low thermal expansion, high density when heavy cations are used (Lu, Gd), high scintillation efficiency and, especially in medical applications, low toxicity. Garnets A$_3$B$_2$C$_3$O$_{12}$ crystallize in a cubic lattice (space group Ia$\bar{3}$d) and contain three distinct crystallographic sites – a large dodecahedral position {A}O$_8$, an octahedral site [B]O$_6$ and the smallest tetrahedral site (C)O$_4$, where A, B and C represent various metal ions or their mixtures [1]. The most efficient known garnet scintillators are composed of smaller rare-earth ions ({A} = Gd, Y, Lu) with aluminum and gallium on [B], (C) sites.

As mentioned in the previous works [9,10,11], the efficiency of scintillation and timing performances (afterglow and slow components) of garnets can be negatively affected by many structural defects occurring during the synthesis process. A so-called anti-site defect (a large ion residing on the octahedral site, e.g. Lu$_{Al}$) is the most important structural defect in garnet scintillators and manifests as an electron trap located closely below the conduction band in

$Lu_3Al_5O_{12}$ [12] The presence of anti-site defect-related traps has been linked to long components in the scintillation decay curve [13]. It was shown that low-temperature synthesis routes lead to a decreased presence of such defects [10,14,15] when compared to a typical grow from the melt [13]. The luminescence properties of garnets can be relatively easily influenced by varying their chemical composition [1,10]. The garnet structure enables easy incorporation of most trivalent lanthanide ions and a large variety of other metal ions, leading to the changes in the band gap, so-called band gap engineering [1,16]. Substituting $Al^{3+}$ in LuAG for $Ga^{3+}$ causes lowering of the conduction band minimum until the trap level corresponding to anti-site defects becomes a part of the conduction band and thus stops negatively influencing the scintillation process [5]. However, this also lowers the energy difference (i.e. a thermal barrier) between $Ce^{3+}$ 5d excited states and the conduction band. Gd doping decreases the position of the $Ce^{3+}$ excited state to prevent thermal ionization connected to this energy difference. The optimal concentration of both ions is necessary to prevent thermal ionization as the negative effect of $Ga^{3+}$ ions [17]. Doping with $Ga^{3+}$ also presents a certain technological obstacle due to the volatility of Ga oxides at high temperatures essential to achieve the phase purity, especially in growth from the melt, so the optimization of the heat treatment temperature is a necessary step to reach as high scintillating effectivity as possible. In the case of melt grown synthesis routes, $Sc^{3+}$ can be used instead of $Ga^{3+}$ as it plays the same role and features a lower volatility [18].

In addition to elimination of long decay components through band gap engineering, a further suppression of slower components in decays and increased light yields is possible with co-doping with $Ce^{3+}$ and either $Mg^{2+}$ or $Ca^{2+}$ substituted at a trivalent cation site, as already shown in [3,19]. This aliovalent doping was shown to increase the concentration of $Ce^{4+}$ states and was reported to cause a positive contribution to the fast scintillation response [19,20]. A faster response of the detection system will enable better spatiotemporal resolution of the detector, which would be advantageous, for example, for fast radiography. Although single crystals with perfect transparency can be grown, they suffer from segregation of elements during crystallization from the melt, especially when the ions incorporated into the same crystallographic position have different diameters. Wet chemical methods bypass the segregation from the melt and may provide better defined samples with known composition.

In this paper, the multi-component Ce,Mg-co-doped $Lu_{2.5}Gd_{0.5}Ga_2Al_3O_{12}$ (LuGGAG) with optimized concentration of Ga and Gd was synthesized and the effect of Ce (0.1-1 mol% in irradiated solutions) and Mg (0-10 mol% in irradiated solutions) concentration on its luminescence properties was studied. Based on our previous experiments [9,10], material with this composition provides a high $Z_{eff}$, is phase-pure, and at the same time shows very high radioluminescence intensities. The room-temperature photo-induced synthesis [9] of the LuGGAG solid precursor was used, followed by heat treatment at 1200°C in air. In this study, we focused mostly on the scintillating properties to find the optimal $Ce^{3+}$ and $Mg^{2+}$ concentrations to reach good radioluminescence (RL) intensity and suppress the slower component in decays. In addition, we tried to provide a first quantitative estimate of the Mg within the solid phase using laser-induced breakdown spectroscopy.

## 2. Materials and methods

*Synthesis*

Multicomponent garnets were synthesized by photo-induced homogeneous precipitation already described in [9,10], followed by heat treatment. Aqueous solutions containing 2.5 mmol·dm$^{-3}$ $Lu(NO_3)_3$, 0.5 mmol·dm$^{-3}$ $Gd(NO_3)_3$, 3 mmol·dm$^{-3}$ $Al(NO_3)_3$, 2 mmol·dm$^{-3}$ $Ga(NO_3)_3$ and 0.1 mol·dm$^{-3}$ ammonium formate ($HCOONH_4$) were used for the synthesis. Stock aqueous solutions of cerium nitrate hexahydrate ($Ce(NO_3)_3·6H_2O$) and magnesium nitrate hexahydrate ($Mg(NO_3)_2·6 H_2O$) were used for doping. All chemicals were purchased from Merck with a purity of 99.999% (trace metals basis). The Ce doping used was 0.1 to 1 mol. % relative to the rare-earth elements (RE). As we found out in an earlier study [10], magnesium does not precipitate in solid phase quantitatively, therefore relatively high Mg concentrations in the irradiated solution were chosen and the real Mg concentration in the solid phase was then evaluated by the LIBS method. Samples were consistently labeled with Mg concentration in input solutions. The amount of Mg used in the solution corresponds to the range of 1-10 mol. % relative to the rare-earth elements. For the study on Mg,Ce-codoped samples, the Ce concentration was fixed at 1%; at this concentration, the maximum RL intensity was observed in Ce-doped samples. Solutions were irradiated for 4 hours under stirring by immersible low-pressure mercury lamps (Philips TUV 25W 4P SE) with total power input 100 W. The irradiation caused the formation of fine gelatinous product, which was then separated via microfiltration (0.45 μm filter) and dried at 40°C. Heat treatment of solid phase at 1200°C in air was performed in a Clasic 0415 VAC vacuum furnace.

*Characterization methods*

X-ray powder diffraction (XRPD) for structure characterization and phase purity confirmation was measured using Rigaku MiniFlex 600 diffractometer equipped with copper X-ray tube (Cu-K$_{\alpha 1,2}$ radiation, λ = 1.54184 Å; 40 kV, 15 mA), Ni filter, variable divergence slit below 20° 2θ and NaI:Tl scintillation detector. The measurement was performed in a continuous mode in the range of 10°–80° 2θ with collection speed 2°/min; the width of the data collection interval was 0.02°. Measured data were evaluated in the PDXL2 program using International Center for Diffraction Data (ICDD) PDF-2 database, version 2013. The lattice parameters of the garnet phase as well as crystallite size were determined from the observed diffraction peaks positions and integral width, respectively.

Laser-induced breakdown spectroscopy of pelletized samples was performed using a tunable laser Ekspla NT342C (OPO based, 210-2600 nm) set up at 355 nm, 15 mJ, pulse length 5 ns. Plasma emission spectra were recorded by a LOT-Oriel MS257 spectrometer (grating 2000 lines/mm) coupled with image intensifier (photocathode, microchannel plate) and iCCD detector Andor DH720i-18F-03. Timing parameters were: gate step 2.4 ns, gate width 1000 ns, gate delay 500 ns.

Luminescence properties were evaluated by measuring room-temperature radioluminescence emission spectra (RL) using the custom-made spectrofluorometer 5000M (Horiba Jobin Yvon), equipped with the X-ray source DEBYFLEX ID3003 (Seifert GmbH) with tube DX-W 10x1-S 2400 W (tungsten anode). Detection part of the used setup consists of a single grating monochromator and photon counting detector TBX-04 (IBH Scotland).

Pulsed X-rays with energies up to 30 keV were generated with a repetition rate of 100 kHz by a picosecond diode laser at 405 nm (Delta diode from Horiba) focused on an X-ray tube (model N5084 from Hamamatsu). The resulting photons were detected by a hybrid PMT 140-C from Becker & Hickl GmbH. For decay-time measurements, the photons were histogramed using a PicoHarp300 time-correlated single-photon counting (32 ps time/ bin) and for the time resolved spectra, a MCS6A multiple-channel time analyzer was used (800 ps time/bin). The powders were placed in a custom sample holder to obtain the relative scintillation yields.

## 3.  Results and discussion

### 3.1 Structural analysis

The structural analysis by XRPD (a typical diffraction pattern is shown in Fig. 1) confirmed that all products after heat treatment consisted of pure garnet phase without any impurities. Lattice parameters *a* slightly increase with the Ce concentration (see Tab. 1) due to the incorporation of large Ce$^{3+}$ ions. However, the Ce content was relatively low and its effect on the lattice parameter was comparable to experimental uncertainties; therefore, no clear linear trend was observed (see Fig. 1 inset). The expected value of *a* for undoped Lu$_{2.5}$Gd$_{0.5}$Ga$_2$Al$_3$O$_{12}$ based on the assumption of Vegard's rule validity is approx. 12.051 Å [9]; the experimental data for 0.1% Ce sample match this value within experimental uncertainty. On the other hand, clear increases in the lattice parameters with increasing Mg concentration in the solutions were observed (see Tab. 2); the linear increase in *a* may correspond to the incorporation of large $Mg^{2+}$ ions into the octahedral sites of the garnet lattice. As $Mg^{2+}$ can be expected to enter both the dodecahedral and octahedral sites, the values of *a* cannot easily be used to estimate Mg doping level. The crystallite size of all samples (see Tab. 1, Tab. 2) varies in all samples in the range of higher tens of nm, with no discernible correlation to composition.

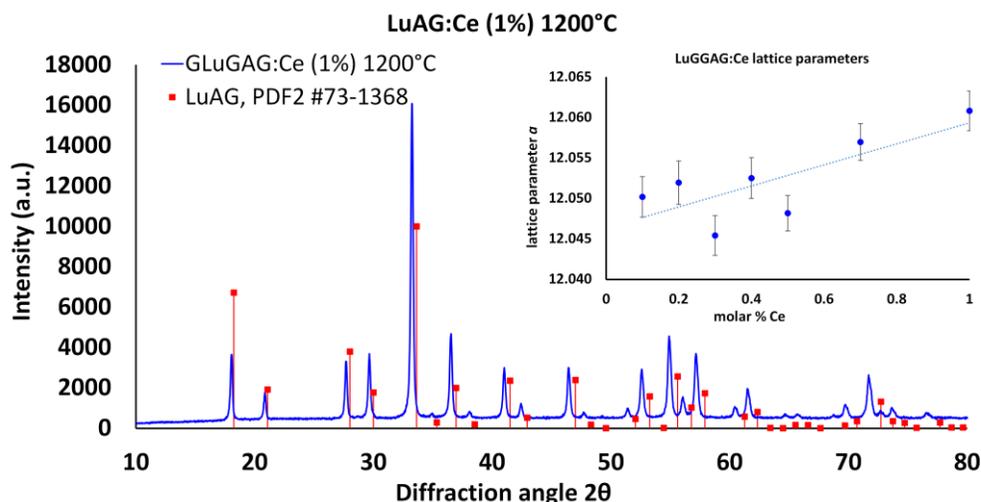

Fig. 1: X-ray diffraction pattern of the LuGGAG:1% Ce sample; inset: determined lattice parameters of samples with different Ce dopation

Tab. 1: Lattice parameters and linear crystallite size of LuGGAG:Ce samples for various Ce doping, calculated from XRD data

| Ce [mol. %] | *a* [Å]    | size [nm] |
|-------------|------------|-----------|
| 0.1         | 12.050(2)  | 72        |
| 0.2         | 12.052(3)  | 73        |
| 0.3         | 12.045(2)  | 74        |
| 0.4         | 12.053(3)  | 86        |
| 0.5         | 12.048(2)  | 80        |
| 0.7         | 12.057(2)  | 60        |
| 1.0         | 12.061(2)  | 74        |

Tab. 2: Lattice parameters and linear crystallite size of LuGGAG:Mg,Ce samples for various Mg content in the solutions (1% Ce doping), calculated from XRD data, and the real Mg content estimated by the LIBS method

| Mg [mol. %] | *a* [Å]   | size [nm] | Mg content by LIBS [mol. %] |
|-------------|-----------|-----------|------------------------------|
| 1           | 12.046(2) | 95        | 0.8                          |
| 2 | 12.053(3) | 68 | 1.1 |
| 4 | 12.051(3) | 62 | 1.1 |
| 6 | 12.078(5) | 42 | 3.9 |
| 8 | 12.095(6) | 33 | 4.1 |
| 10 | 12.099(2) | 61 | 5.0 |

*3.2 Elemental analysis*

The XRF spectra of synthesized solid precursors confirmed the presence of all constituent elements (except Mg); typical spectra are shown in Fig. S4 and the relevant analytical lines for LuGGAG:Ce (0.1–1%) are summarized in Tab. S2. Due to the complex matrix, calibration samples were not feasible; however, the Ce content in the solid phase increases linearly with its concentration in the initial solutions. After each synthesis, the presence of free lanthanides in the solutions was tested by potassium oxalate; as no precipitate was observed, we assume Ce concentration to be nominal.

Mg evaluation relied on the spectroscopic measurements of laser-generated plasma in LIBS, as the XRF cannot detect Mg in small or moderate quantities. The typical spectrum shown in Fig. S5 features numerous spectral lines in the UV range (270–305 nm), including the elemental $Mg^0$ emission (Mg I) at 285.3 nm and ionized $Mg^+$ emission (Mg II) at 280.3 nm. It was observed that for identical experimental setting, the intensity ratio Mg I / Mg II decreases with rising Mg content in the samples, probably due to changes in ionization efficiency and plasma temperature. The ratio between the Mg I and Lu II lines at 284.7 nm was used to estimate Mg/Lu ratio in the samples; several smaller Mg I lines overlap with this Lu II line, but their contribution was found negligible at low Mg contents in the present samples. The Mg content in solid phase was found to increase linearly with the concentration of Mg in the initial solution, with some outliers (see Fig. 2). The outliers may be either caused by synthesis (an inconsistent co-precipitation of Mg) or the damage to pellets and their disintegration during LIBS measurements. After calibration, it accounted to ~0.8–5.0 % (see Tab. 2); the decrease with respect to concentration in the solutions was probably caused by the mild solubility of Mg compounds in aqueous solutions.

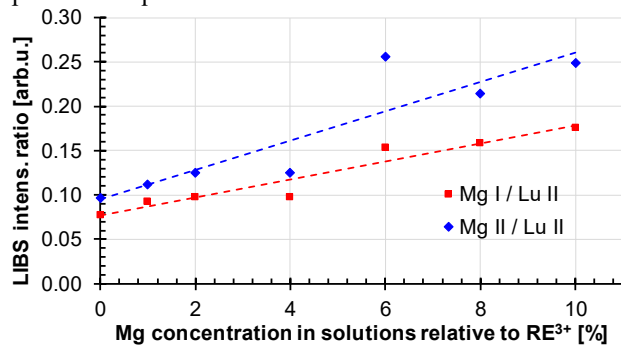

Fig. 2: Comparison between LIBS spectral intensities of Mg and the concentrations of $Mg^{2+}$ in the solution used for synthesis

*3.3 Luminescence properties*

All prepared samples with various concentration of $Ce^{3+}$ or co-doped with various concentration of $Mg^{2+}$ were characterized under X-rays excitation. Radioluminescence emission spectra of LuGGAG:Ce (0.1–1%) are shown in Fig. 3; the luminescence intensity can be quantitatively compared. The observed wide emission band between 480 nm and 550 nm is typical for the transition from the excited $5d^1$ state of $Ce^{3+}$ to its ground state $4f^1$ doublet ($^2F_{5/2}$, $^2F_{7/2}$); its intensity increased with rising $Ce^{3+}$ concentration (see Fig. 4). A weak emission line in UV area at 312 nm can be ascribed to the 4f-4f $Gd^{3+}$ transition; its intensity decreases with increasing $Ce^{3+}$ concentration (see Fig. 4). This decrease may be attributed to either energy transfer from $Gd^{3+}$ to $Ce^{3+}$, or competition for charge carriers. The highest intensity of RL emission was observed for doping with 1% of $Ce^{3+}$, so this composition was used for study of Mg doping effect.

It is pertinent to mention here that the maximal obtained RL emission peak area is 6 times higher than the RL emission area of BGO, which was used as a reference scintillating material.

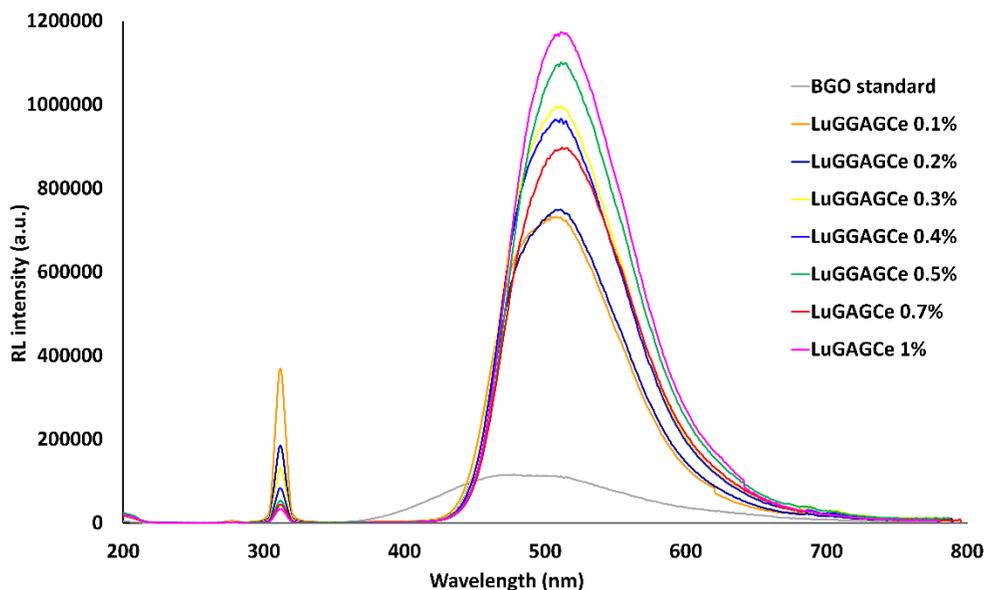

Fig. 3: Steady-state radioluminescence spectra of LuGGAG:Ce samples

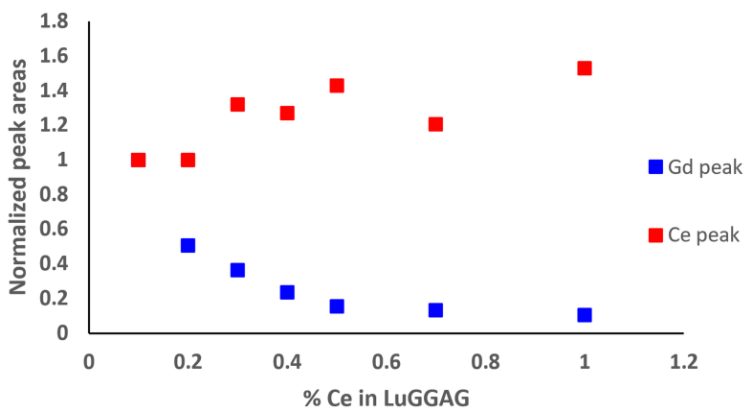

Fig. 4: RL emission peak areas of LuGGAG:Ce samples normalized to the 0.1% Ce sample

RL spectra of LuGGAG:Ce,Mg with various concentrations of Mg in irradiated solutions are shown in Fig. 5, featuring both Ce and Gd emission bands with identical shape in all samples, but with varying intensity (normalized spectra are shown in Fig. S1 and S2). It is evident that the RL intensity decreases with increasing Mg content (see Fig. 6). This effect can be explained by an excessive creation of $Ce^{4+}$ states; the mechanism was already described in [3] and experimentally confirmed for example in [10], where an indirect determination via the ratios of the photoluminescence excitation band intensities was described. $Ce^{4+}$ has a strong absorption band in the UV region, which absorbs light before it can excite $Ce^{3+}$ through the

4f-5d$_2$ transition at 340 nm. In the presence of Ce$^{4+}$, this PLE band thus decreases in intensity compared to the 4f-5d$_1$ band at ~ 440 nm. Regardless of luminescence intensity changes, formation of Ce$^{4+}$ through Mg$^{2+}$ co-doping and charge compensation leads to the suppression of slower components in PL and RL decays [19], as shown in Fig. 7 and supplementary information Fig. S3 and Tab. S1. Scintillation decay curves were fitted by a sum of three exponential functions. All decay components of RL decays decrease sharply with increasing Mg concentration (see Fig. 7 b,c,d). The mean decay time decreases as well (see Fig. 7 e). The relative light yield was evaluated from scintillation decay curve as well (see Fig. 7 f) within a 50 ns and 8 µs time window, respectively. Both aspects (acceleration of decays and LY decrease in highly Mg-doped samples) have been observed in Ce,Mg co-doped epitaxial thin films of aluminate garnets as well [20]. Weak Gd$^{3+}$ emission at 312 nm is still evident and this emission diminishes with increasing amount of Mg to a very similar degree to Ce$^{3+}$ emission decrease (see Fig. 6). This might point either to a Mg-related defect in the samples that competes for charge carriers with both Gd and Ce, or to the absorption of this emission line within the Ce$^{4+}$-O$^{2-}$ charge transfer band in UV. Absolute RL emission intensity of samples co-doped with Mg (LuGGAG:Ce,Mg) is lower in comparison with samples doped only with 1% of Ce$^{3+}$ (LuGGAG:Ce (1%)). Evaluating the optimal composition is quite difficult; however, in terms of the number of emitted photons within the first 3 ns, the sample with 1% Ce and the real Mg content in the solid phase of 2–4% appears to be the best in our investigated area. The calculated percentage of photons in the first 3 ns is 6.9 % for 1% Mg (in solution), while it rises to 19 % for 8% Mg doping (in solution).

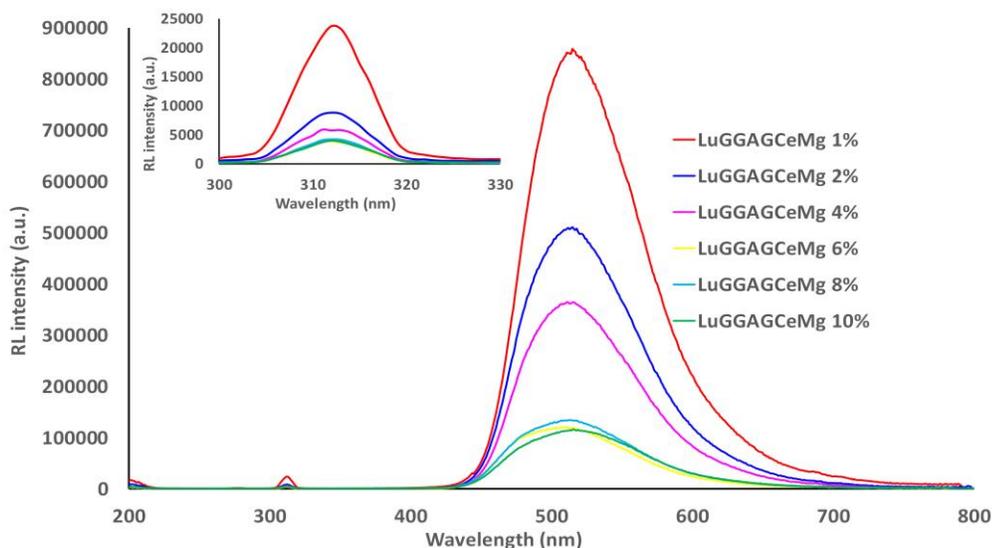

Fig. 5: Steady-state radioluminescence spectra of LuGGAG:Ce(1%),Mg samples

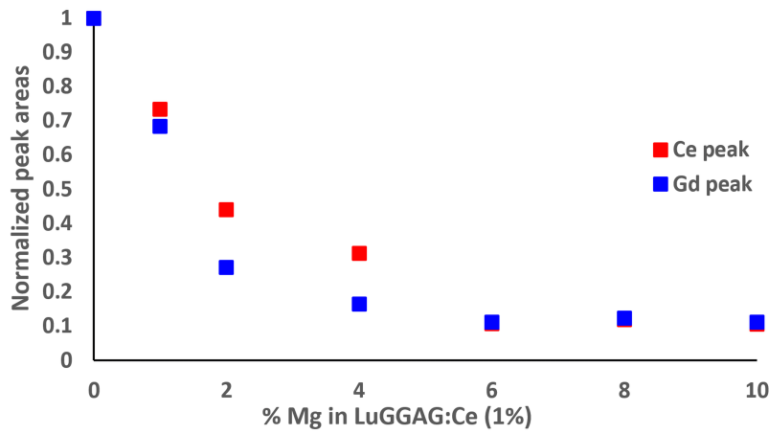

Fig. 6: RL emission peak areas of LuGGAG:Ce,Mg samples normalized to the 1% Ce sample

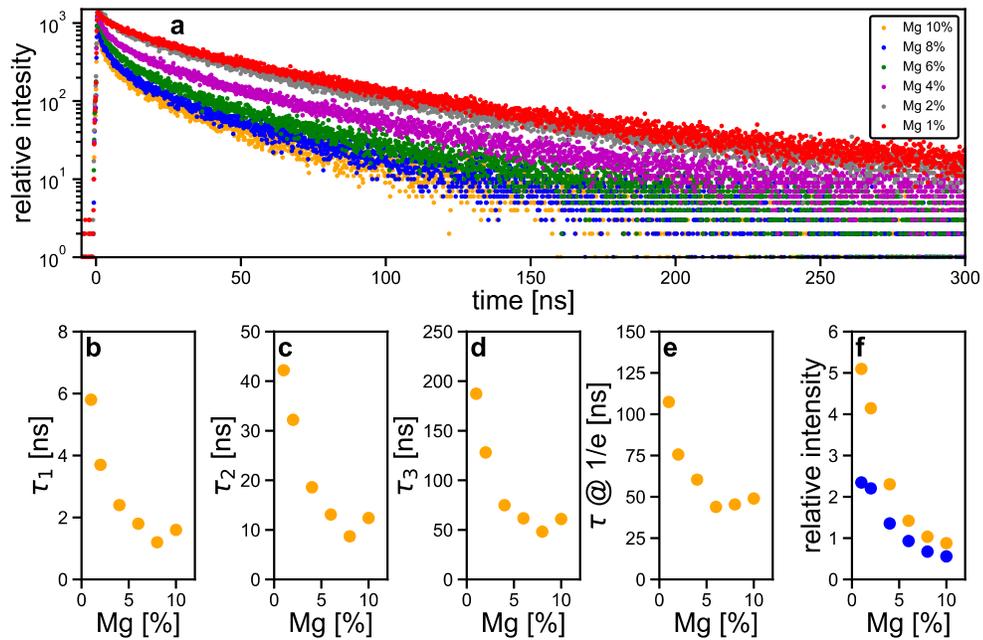

Fig. 7: -a- Scintillation decay time under X-ray excitation recorded at 530 ±10 nm, -b,c,d- Evolution of the 3 decay time fit components as a function of the Mg concentration. -e- Evolution of the time for 1/e fraction of the emitted photons, -f- Evolution of the relative yield over the whole recorded time (orange; 8 μs, excited at 100 kHz) and in the first 50 ns (blue)

## 4. Conclusions

Photo-induced synthesis was shown to be a convenient method for the synthesis of nanocrystalline multicomponent LuGGAG garnets doped with $Ce^{3+}$ and $Mg^{2+}$ ions in a relatively wide concentration range. LuGGAG:Ce (0.1 to 1 % Ce) and LuGGAG:Ce,Mg (1%

Ce, Mg concentration varying in the range of 1 to 10 %) series were synthesized and their phase composition was confirmed; the Ce doping caused a rather small increase in lattice parameter, while the Mg content led to a marked increase reasonably linear with respect to initial $Mg^{2+}$ concentration in the solution. Ce concentration in the solid phase was confirmed by XRF results, while the Mg was shown by LIBS to have a slightly lower concentration than the composition of initial solutions. Results show that increasing Ce concentration leads to both a strong increase in luminescence intensity attributed to the $Ce^{3+}$ emission centers, and a decrease in Gd emission peak intensity. Optimal value of Ce doping was established as 1 %; further increase in Ce concentration would lead to concentration quenching. Co-doping with Mg ions led to a decrease in steady-state radioluminescence intensity of $Ce^{3+}$ emission, as well as accelerating of PL and RL decays and a significant reduction of the slow component in RL decays. In terms of the fraction of emitted photons in the first 3 ns, the sample prepared from a solution with 8% Mg with a simultaneous doping of 1% Ce, appears to be the best in our investigated area.

## 5. Acknowledgments

This research has been supported by the Ministry of Education, Youth and Sports of the Czech Republic, project SENDISO, grant number CZ.02.01.01/00/22_008/0004596, the Czech Science Foundation, grant number GA23-05615S, and by the European Community through the grant number 899293, HORIZON 2020 – SPARTE.